\documentclass[11pt,prd,onecolumn,amsmath,amssymb,aps,floats,floatfix,nofootinbib]{revtex4-2}
\usepackage[colorlinks=true,urlcolor=blue,anchorcolor=blue,citecolor=blue,filecolor=blue,linkcolor=blue,menucolor=blue,linktocpage=true]{hyperref} 

\usepackage{setspace}
\usepackage{graphicx} 
\usepackage{lineno}
\usepackage{color,ulem,makecell}
\usepackage{graphicx} 
\usepackage{subcaption} 
\usepackage{slashed}
\usepackage{multirow}
\usepackage{makecell}
\usepackage{booktabs}
\usepackage{caption}
\usepackage{xcolor}
\usepackage{tabularx}
\usepackage{amsmath}
\usepackage[left=2.5cm,right=2.5cm,bottom=3cm]{geometry}
\usepackage[utf8]{inputenc} 
\usepackage[scientific-notation=true]{siunitx} 

\graphicspath{{figure/}}

\allowdisplaybreaks 

\newcolumntype{C}{>{\centering\arraybackslash}X}
\newcolumntype{L}{>{\raggedright\arraybackslash}X}

\begin{document}

\title{Machine learning study on single production of a singlet vectorlike lepton at the Large Hadron Collider}

\author{Yiheng Cui}
\author{Shiyu Wang}
\author{Zhao-Huan Yu}\email[Corresponding author. ]{yuzhaoh5@mail.sysu.edu.cn}
\author{Hong-Hao Zhang}\email[Corresponding author. ]{zhh98@mail.sysu.edu.cn}
\affiliation{School of Physics, Sun Yat-Sen University, Guangzhou 510275, China}

\begin{abstract}
Vectorlike leptons are nonchiral, colorless fermions from new physics beyond the Standard Model, appearing in many theoretical extensions.
We investigate the prospect for detecting the single production of a singlet vectorlike lepton that mixes with the $\tau$ lepton at the Large Hadron Collider.
The corresponding final states are classified as the three- and four-lepton search channels.
The machine learning algorithm XGBoost is employed to enhance signal-background discrimination.
Our analysis indicates that, at $\sqrt{s} = 14~\mathrm{TeV}$ with an integrated luminosity of $3000~\mathrm{fb}^{-1}$ under the assumption of negligible systematic uncertainties, the expected $2\sigma$ exclusion limits in the three- and four-lepton channels can reach vectorlike lepton masses up to $500$ and $405~\mathrm{GeV}$ in the parameter region allowed by the electroweak oblique parameter constraint, respectively.
These findings demonstrate that machine learning techniques can substantially improve the sensitivity of collider searches for vectorlike leptons.
\end{abstract}

\maketitle
\tableofcontents

\section{Introduction}
\label{sec:intro}

Several extensions of the Standard Model (SM) predict new heavy leptons, particularly vectorlike leptons (VLLs), which transform as either $\mathrm{SU(2)_L} $ singlets or doublets.
Such VLLs naturally arise in various frameworks beyond the SM, including left-right symmetric models~\cite{Mohapatra:1974gc,Pati:1974yy,Mohapatra:1977mj,Bahrami:2016has}, grand unified theories~\cite{Morais:2020ypd,Morais:2020odg,Li:2010hi,Quinta:2025kal}, dark matter models~\cite{FileviezPerez:2013eoz,Schwaller:2013hqa,Halverson:2014nwa,Yu:2014mfa,Bhattacharya:2021ltd}, composite models~\cite{Cacciapaglia:2022zwt,Cacciapaglia:2020kgq}, and supersymmetric models~\cite{Martin:2009bg,Araz:2018uyi,Endo:2011mc,Kitano:2000zw,Joglekar:2013zya}, as well as models~\cite{Das:2020gnt,Das:2020uer,Cherchiglia:2021syq,Kawamura:2022fhm,Escribano:2021css,Antusch:2023mqe,Kawamura:2023zuo} based on the type-I~\cite{Minkowski:1977sc,Gell-Mann:1979vob,Yanagida:1979as} and type-III~\cite{Foot:1988aq} seesaw mechanisms.  
Therefore, searches for VLLs serve as an important avenue for exploring new physics.

The Large Hadron Collider (LHC) has searched for physics beyond the SM in proton-proton collisions at center-of-mass energies up to  $ \sqrt{s} = 13.6~\mathrm{TeV}$.  
Following the discovery of the Higgs boson with a mass of  $ m_h \approx 125~\mathrm{GeV}$ in 2012, current data are consistent with the SM, and both direct and indirect constraints on new physics now extend to the TeV scale.
As the mass scale of new physics increases, many models exhibit decoupling behaviors~\cite{Bhattiprolu:2019vdu}.  
New fermions whose masses arise primarily from bare electroweak singlet terms in the Lagrangian,
rather than from Yukawa couplings to the Higgs field,
can naturally evade stringent experimental bounds and remain theoretically viable.
VLLs are prime examples of such fermions and can address specific theoretical challenges in new physics models.
For instance, in weak-scale supersymmetry, the mass of the lightest Higgs boson can be raised to meet the observation by introducing new heavy chiral supermultiplets containing VLLs with large Yukawa couplings~\cite{Martin:2010dc}.

In this work, we focus on an $\mathrm{SU(2)_L}$ singlet VLL $\tau^\prime$, which carries the same lepton number as the $\tau$ lepton.
Consequently, $\tau'$ decays predominantly into final states containing a $\tau $ lepton.
The CMS Collaboration has reported searches for VLLs in the  $\tau^\prime \to E\nu$  channel ($ E = e, \mu, \tau $) at the $13~\mathrm{TeV} $ LHC with an integrated luminosity of $138~\si{fb^{-1}}$ and excluded a singlet VLL with a mass $m_{\tau^\prime}$ between $\sim 125$ and $\sim 170~\mathrm{GeV}$ at the $95\%$ confidence level (CL)~\cite{CMS:2024bni}.
In addition, a previous study examined the sensitivity of LHC searches at $\sqrt{s} = 8~\mathrm{TeV}$ and $ 13~\mathrm{TeV} $ to $\tau'$ using multilepton final states~\cite{Kumar:2015tna}.
It was found that for the four-lepton channel at the 13~TeV LHC with $1000~\mathrm{fb}^{-1} $ of integrated luminosity the $\tau'$ mass $m_{\tau'}$ is expected to be excluded up to $200~\mathrm{GeV} $ at the $95\%$ CL.
These results indicate that traditional LHC searches for a singlet VLL have rather limited sensitivity.
Nevertheless, if $\tau'$ decays into a long-lived light pseudoscalar boson and a $\tau$ lepton, the $95\%$ CL exclusion limit can be extended to $m_{\tau'} \sim 700~\mathrm{GeV} $, depending on the lifetime of the pseudoscalar, as suggested by the CMS analysis with a dataset of $138~\si{fb^{-1}}$ at $\sqrt{s} = 13~\mathrm{TeV}$~\cite{CMS:2025urb}.

Machine learning techniques offer significant potential to enhance the sensitivity of collider searches. For instance, one study has demonstrated that the high-luminosity LHC (HL-LHC), when integrated with machine learning methods, can achieve sensitivity to $\mathrm{SU(2)_L}$ doublet VLLs with masses up to the TeV scale~\cite{Freitas:2020ttd}.
Therefore, in this work, we employ machine learning techniques to investigate how the LHC sensitivity to a singlet VLL can be improved.
The signal process considered involves the production of a singlet VLL $\tau'$ in association with  a $\tau$ lepton in proton-proton collisions.
Notably, this $\tau$ lepton originates directly from the hard interaction alongside the $\tau'$ lepton, rather than from its subsequent decay, which induces another $\tau$ lepton.

Depending on the decay modes of the two $\tau$ leptons, the final states can be classified into a three-lepton channel, where one $\tau$ lepton decays hadronically and the other decays leptonically, and a four-lepton channel with two leptonically decaying $\tau$ leptons.
In order to enhance the discrimination between signal and background events, the machine learning algorithm XGBoost is employed, using several kinematic observables as input features.
The primary observables include the transverse momentum of the leading lepton, the missing transverse energy, the scalar sum of the transverse momenta of two leading leptons, and the invariant masses of two-, three-, and four-lepton systems.
The sensitivity of the HL-LHC at $\sqrt{s} = 14~\mathrm{TeV}$ is subsequently investigated.

The paper is organized as follows. Section~\ref{sec:model} presents the theoretical framework of the singlet VLL model. 
Section~\ref{sec:oblique} investigates the constraint on the model parameters from electroweak oblique parameters.
Section~\ref{sec:simulation} discusses the Monte Carlo simulation setup for LHC searches and the baseline analysis results for the three- and four-lepton search channels using simulated samples. In Sec.~\ref{sec:ml}, the XGBoost algorithm is applied to further optimize the signal-background discrimination, and the sensitivity of HL-LHC is estimated. Finally, Sec.~\ref{sec:conclusion} provides a summary and discusses the implications of the findings.

\section{Simplified Model}
\label{sec:model}

We consider a simplified model, where the SM is extended by only a vectorlike lepton $ \tau^{\prime 0} $, which is a colorless $ \mathrm{SU(2)_L} $  singlet with weak hypercharge $Y = -1$ and electric charge  $ Q = -1 $. 
The superscript ``0'' denotes that the $ \tau^{\prime 0} $ field is a gauge eigenstate.
Following a parametrization similar to that used for vectorlike quarks~\cite{Buchkremer:2013bha,Aguilar-Saavedra:2013qpa}, we adopt a simple framework that captures the main phenomenological features of the VLL $ \tau^{\prime 0} $.
The related terms in the Lagrangian can be written as
\begin{equation}
    \begin{aligned}
        \mathcal{L} \supset i \bar{\tau}^{\prime 0} \gamma^{\mu} D_{\mu} \tau^{\prime 0} - (M \, \bar{\tau}^{\prime 0}_{L} \tau^{\prime 0}_{R}
+ \varepsilon \, \bar{L} H \tau^{\prime 0}_{R}
+ y_{\tau} \, \bar{L} H \tau^{0}_{R}
+ \text{H.c.}),
    \end{aligned}
\end{equation}
where $\tau^0$ is the gauge eigenstate of the $\tau$ lepton field, and
\begin{equation}
    \begin{aligned}
        H = \frac{1}{\sqrt{2}}\begin{pmatrix} 0 \\ v + h\end{pmatrix},
\qquad
L = \begin{pmatrix} \nu_{\tau L} \\ \tau^{0}_L 
\end{pmatrix}
    \end{aligned}
\end{equation}
are the SM Higgs doublet in the unitary gauge and the third-generation lepton doublet, respectively.
$v = 246.22~\si{GeV}$ is the Higgs vacuum expectation value, and $h$ is the Higgs boson.
$\nu_\tau$ is the $\tau$ neutrino.
The covariant derivative of $\tau^{\prime 0}$ is
\begin{equation}
  D_{\mu} \tau^{\prime 0} = (\partial_{\mu} + i g' B_{\mu}) \tau^{\prime 0} =
  \partial_{\mu}\tau^{\prime 0} + i g' (c_W A_{\mu} - s_W Z_{\mu}) \tau^{\prime 0},
\end{equation}
where $B_\mu$, $A_\mu$, and $Z_\mu$ are the $\mathrm{U(1)_Y}$ gauge field, the photon field, and the $Z$ gauge field, respectively.
$c_W \equiv \cos \theta_W$ and $s_W \equiv \sin \theta_W$ are two trigonometric functions of the weak mixing angle $\theta_W$.

We assume that the mass parameter $M$ and the coupling constants $y_{\tau}$ and $\varepsilon$ are all real.
After the electroweak symmetry breaking, the mass terms become
\begin{equation}
 \mathcal{L}_\mathrm{mass} = 
- \begin{pmatrix} \bar{\tau}^{0}_{L} & \bar{\tau}^{\prime 0}_{L} \end{pmatrix}
\begin{pmatrix} y_{\tau} v/\sqrt{2} & \varepsilon v/\sqrt{2} \\ 0 & M \end{pmatrix}
\begin{pmatrix} \tau^{0}_{R} \\ \tau^{\prime 0}_{R} \end{pmatrix}
+ \text{H.c.}
\end{equation}
In order to diagonalize the mass matrix, we introduce two real orthogonal matrices
\begin{equation}
V_{L,R} = \begin{pmatrix} \cos\theta_{L,R} & -\sin\theta_{L,R} \\ \sin\theta_{L,R} & \cos\theta_{L,R} \end{pmatrix},
\end{equation}
where $\theta_L$ and $\theta_R$ are the left- and right-handed $\tau$-$\tau'$ mixing angles, respectively.
The mass terms can be rewritten as
\begin{eqnarray}
\mathcal{L}_\mathrm{mass} &=& 
- \begin{pmatrix} \bar{\tau}^{0}_{L} & \bar{\tau}^{\prime 0}_{L} \end{pmatrix}
V_L V_L^\mathrm{T} \begin{pmatrix} y_{\tau} v/\sqrt{2} & \varepsilon v/\sqrt{2} \\ 0 & M \end{pmatrix} V_R V_R^\mathrm{T}
\begin{pmatrix} \tau^{0}_{R} \\ \tau^{\prime 0}_{R} \end{pmatrix}
+ \text{H.c.}
\nonumber\\
 &=& - m_{\tau} \bar{\tau}_{L} \tau_{R}  -  m_{\tau'} \bar{\tau}'_{L} \tau'_{R}  + \text{H.c.},
\end{eqnarray}
where the mass eigenstates are defined by
\begin{equation}
\begin{pmatrix}
\tau_{L} \\
\tau^{\prime}_{L}
\end{pmatrix}
\equiv
V_L^\mathrm{T}
\begin{pmatrix}
\tau^{0}_{L} \\
\tau^{\prime 0}_{L}
\end{pmatrix},\quad
\begin{pmatrix}
\tau_{R} \\
\tau^{\prime}_{R}
\end{pmatrix}
\equiv
V_R^\mathrm{T}
\begin{pmatrix}
\tau^{0}_{R} \\
\tau^{\prime 0}_{R}
\end{pmatrix}.
\end{equation}
The diagonalization relation
\begin{equation}
V_L^\mathrm{T}
\begin{pmatrix}
y_{\tau} v/\sqrt{2} & \varepsilon v/\sqrt{2} \\
0 & M
\end{pmatrix}
V_R
= 
\begin{pmatrix}
m_\tau & 0 \\
0 & m_{\tau'}
\end{pmatrix}
\end{equation}
leads to
\begin{alignat}{2}
\tan\theta_R &=\frac{m_{\tau}}{m_{\tau'}} \tan\theta_L,&\quad
M &= m_{\tau} s_L s_R + m_{\tau'} c_L c_R,
\\
y_{\tau} &= \frac{\sqrt{2}}{v}  (m_{\tau} c_L c_R + m_{\tau'} s_L s_R),&\quad
\varepsilon &= \frac{\sqrt{2}}{v} (m_{\tau} c_L s_R - m_{\tau'} s_L c_R),
\end{alignat}
where the shorthand notations $c_{L,R} \equiv \cos\theta_{L,R}$ and $s_{L,R} \equiv \sin\theta_{L,R}$ are used.
Moreover, the left-handed $\tau$-$\tau'$ mixing angle $\theta_L$ satisfies
\begin{equation}
  \tan 2 \theta_L = \frac{\sqrt{2} \varepsilon v M}{M^2 - (y_{\tau}^2 v^2 +
  \varepsilon^2 v^2)/2}.
\end{equation}

The $\tau$-$\tau'$ mixing arises from the Yukawa coupling $\varepsilon$.
Therefore, $\varepsilon = 0$ leads to $\theta_L = \theta_R = 0$.
The mass of the $\tau$ lepton is fixed by its observational value $m_\tau = 1776.93 \pm 0.09~\si{MeV}$~\cite{ParticleDataGroup:2024cfk}.
Thus, two free parameters of the model can be chosen as the $\tau'$ mass, $m_{\tau'}$, and the sine of the left-handed $\tau$-$\tau'$ mixing angle, $s_L$.

Expressing in the mass eigenstates, the Yukawa couplings for the $\tau$ and $\tau'$ leptons are given by
\begin{equation}
\mathcal{L}_{\mathrm{Yukawa}} = - \frac{m_{\tau} c_L^2}{v}\, h \bar{\tau}_L \tau_R - \frac{m_{\tau'} s_L^2}{v}\,
h \bar{\tau}_L' \tau_R' + \frac{m_{\tau'} s_L c_L}{v}\, h \bar{\tau}_L \tau_R' +
\frac{m_{\tau} s_L c_L}{v}\, h \bar{\tau}_L' \tau_R + \mathrm{H.c.},
\end{equation}
while the electroweak gauge couplings can be written as
\begin{eqnarray}
\mathcal{L}_{\mathrm{gauge}} &=& - e A_{\mu} (\bar{\tau} \gamma^{\mu} \tau + \bar{\tau}' \gamma^{\mu} \tau') +
\frac{g}{\sqrt{2}} (c_L W^+_{\mu} \bar{\nu}_L \gamma^{\mu} \tau_L - s_L
W^+_{\mu} \bar{\nu}_L \gamma^{\mu} \tau_L' + \mathrm{H.c.})
\nonumber\\
&& + \frac{g}{2 c_W}\, Z_{\mu} \left[ (2 s_W^2 - c_L^2) \bar{\tau}_L \gamma^{\mu}
\tau_L + (2 s_W^2 - s_L^2) \bar{\tau}'_L \gamma^{\mu} \tau_L' + 2 s_W^2
(\bar{\tau}_R \gamma^{\mu} \tau_R + \bar{\tau}'_R \gamma^{\mu} \tau'_R)
\right]
\nonumber\\
&& + \frac{g c_L s_L}{2 c_W}\, Z_{\mu} (\bar{\tau}_L \gamma^{\mu} \tau_L' + \bar{\tau}'_L \gamma^{\mu} \tau_L),
\label{eq:L_gauge}
\end{eqnarray}
where $W^+_\mu$ is the $W$ gauge field, and $e$ and $g$ denotes the $\mathrm{U(1)_{EM}}$ and $\mathrm{SU(2)_L}$ gauge couplings.

\section{Constraints from Electroweak Oblique Parameters}
\label{sec:oblique}

The VLL can modify electroweak precision observables via vacuum-polarization corrections to the electroweak gauge bosons. 
These new physics effects can be parametrized using the electroweak oblique parameters $S$, $T$, and $U$ \cite{Peskin:1990zt,Peskin:1991sw}.
The corresponding one-loop vacuum-polarization functions for the electroweak gauge bosons can be expressed as \cite{Cao:2022mif}
\begin{eqnarray}
	\Sigma_{V'V}(0)
	&=&
	\sum_{i,j}
	\frac{2}{16\pi^2}
	\big[
	(
	g_L^{\bar{\psi}_j\psi_i V'}
	g_L^{\bar{\psi}_i\psi_j V^*}
	+
	g_R^{\bar{\psi}_j\psi_i V'}
	g_R^{\bar{\psi}_i\psi_j V^*}
	)
	F_1(m_{\psi_i},m_{\psi_j})
	\nonumber\\
	&& \qquad
	+
	(
	g_L^{\bar{\psi}_j\psi_i V'}
	g_R^{\bar{\psi}_i\psi_j V^*}
	+
	g_R^{\bar{\psi}_j\psi_i V'}
	g_L^{\bar{\psi}_i\psi_j V^*}
	)
	m_{\psi_i}m_{\psi_j}
	F_2(m_{\psi_i},m_{\psi_j})
	\big],
\\
 	\Sigma'_{V'V}(0) &\equiv& \left.
	\frac{\partial \Sigma_{V'V}(p^2)}
	{\partial p^2}
	\right|_{p^2=0}
	=
	\sum_{i,j}
	\frac{2}{16\pi^2}
	\big[
	(
	g_L^{\bar{\psi}_j\psi_i V'}
	g_L^{\bar{\psi}_i\psi_j V^*}
	+
	g_R^{\bar{\psi}_j\psi_i V'}
	g_R^{\bar{\psi}_i\psi_j V^*}
	)
	F_3(m_{\psi_i},m_{\psi_j})
	\nonumber\\
	&&\qquad
	+
	(
	g_L^{\bar{\psi}_j\psi_i V'}
	g_R^{\bar{\psi}_i\psi_j V^*}
	+
	g_R^{\bar{\psi}_j\psi_i V'}
	g_L^{\bar{\psi}_i\psi_j V^*}
	)
	m_{\psi_i}m_{\psi_j}
	F_4(m_{\psi_i},m_{\psi_j})
	\big],
\end{eqnarray}
where $V,V'=W^\pm,Z,\gamma$ and $\psi_i=\tau',\tau,\nu_\tau$.
Based on the Lagrangian presented in the preceding section, the chiral couplings $g_{L/R}^{\bar{\psi}_j\psi_i V'}$ and $g_{L/R}^{\bar{\psi}_j\psi_i V^*}$ are given by
\begin{align}
&
g_L^{\bar{\tau}\tau\gamma}
=
g_R^{\bar{\tau}\tau\gamma}
=
g_L^{\bar{\tau}'\tau'\gamma}
=
g_R^{\bar{\tau}'\tau'\gamma}
=
-e,
\\[2mm]
&
g_L^{\bar{\nu}\tau W^+}
=
g_L^{\bar{\tau}\nu W^-}
=
\frac{g c_L}{\sqrt{2}},
\qquad
g_R^{\bar{\nu}\tau W^+}
=
g_R^{\bar{\tau}\nu W^-}
=
0,
\\[2mm]
&
g_L^{\bar{\nu}\tau' W^+}
=
g_L^{\bar{\tau}'\nu W^-}
=
-\frac{g s_L}{\sqrt{2}},
\qquad
g_R^{\bar{\nu}\tau' W^+}
=
g_R^{\bar{\tau}'\nu W^-}
=
0,
\\[2mm]
&
g_L^{\bar{\tau}\tau Z}
=
\frac{g}{2c_W}
(
2s_W^2-c_L^2
),
\qquad
g_R^{\bar{\tau}\tau Z}
=
\frac{g s_W^2}{c_W},
\\[2mm]
&
g_L^{\bar{\tau}'\tau' Z}
=
\frac{g}{2c_W}
(
2s_W^2-s_L^2
),
\qquad
g_R^{\bar{\tau}'\tau' Z}
=
\frac{g s_W^2}{c_W},
\\[2mm]
&
g_L^{\bar{\tau}\tau' Z}
=
g_L^{\bar{\tau}'\tau Z}
=
\frac{gc_Ls_L}{2c_W},
\qquad
g_R^{\bar{\tau}\tau' Z}
=
g_R^{\bar{\tau}'\tau Z}
=
0 .
\end{align}
The analytic expressions for the loop functions $F_i(m_{\psi_i},m_{\psi_j})$ are provided in Ref.~\cite{Cao:2022mif}.

Using these vacuum-polarization functions, the electroweak oblique parameters $S$, $T$, and $U$ can be written as
\begin{eqnarray}
	\frac{\alpha}{4s_W^2 c_W^2}\, S
	&\simeq&
	- \Sigma_{ZZ}^{\prime\rm VLL}(0)
	+ \Sigma_{\gamma\gamma}^{\prime\rm VLL}(0)
	+
	\frac{c_W^2-s_W^2}{c_W s_W}\,
    \Sigma_{\gamma Z}^{\prime\rm VLL}(0),
	\\
	\alpha\, T
	&=&
	\frac{\Sigma_{ZZ}^{\rm VLL}(0)}{m_Z^2}
	-
	\frac{\Sigma_{WW}^{\rm VLL}(0)}{m_W^2},
	\\
	\frac{\alpha}{4s_W^2}\, U
	&\simeq&
	- \Sigma_{WW}^{\prime\rm VLL}(0)
	+
	c_W^2 \Sigma_{ZZ}^{\prime\rm VLL}(0)
	+
	s_W^2 \Sigma_{\gamma\gamma}^{\prime\rm VLL}(0)
	+
	2c_W s_W \Sigma_{\gamma Z}^{\prime\rm VLL}.
\end{eqnarray}
Here, $\alpha$ denotes the fine-structure constant, and $m_W$ and $m_Z$ represent the masses of the $W$ and $Z$ bosons.
The VLL contributions are defined as $\Sigma_{V'V}^{\rm VLL}(p^2)=\Sigma_{V'V}(p^2)-\Sigma_{V'V}^{\rm SM}(p^2)$, where $\Sigma_{V'V}^{\rm SM}(p^2)$ corresponds to the contributions within the SM.

\begin{figure}
    \centering
    \includegraphics[width=0.5\linewidth]{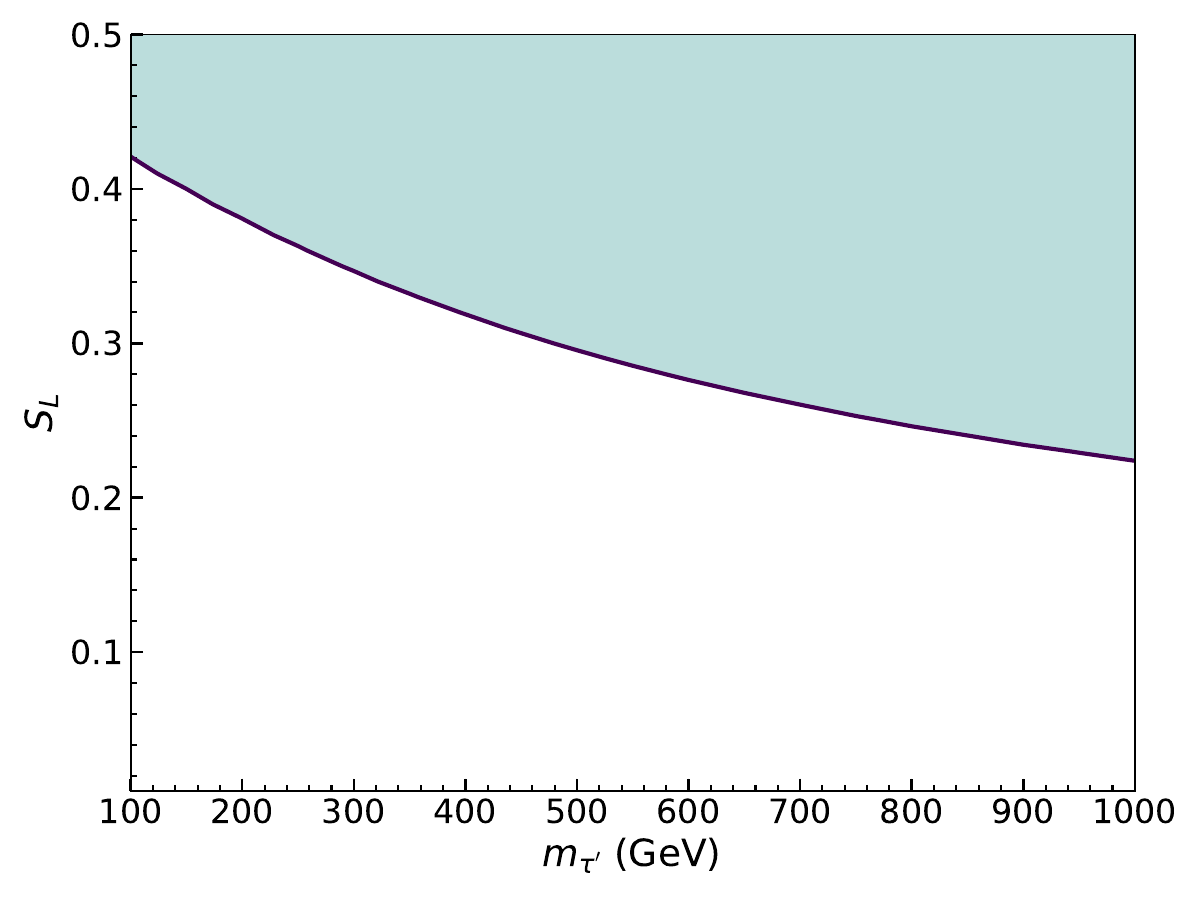}
    \caption{$95\%$ CL exclusion limits from electroweak oblique parameters in the $m_{\tau'}$-$s_L$ plane.}
    \label{fig:chi2}
\end{figure}

Based on the expressions above, we can compute the theoretical predictions of the oblique parameters, $S_{\rm th}$, $T_{\rm th}$, and $U_{\rm th}$, for any chosen values of $m_{\tau'}$ and $s_L$.
The experimental values, taken from the PDG 2025 update of the electroweak precision fit, are~\cite{ParticleDataGroup:2024cfk}
\begin{equation}\label{eq:STU:exp}
    S_{\rm exp}=0.021\pm0.096,\quad
    T_{\rm exp}=0.040\pm0.120,\quad
    U_{\rm exp}=0.008\pm0.092,
\end{equation}
with correlation coefficients $\rho_{ST} = 0.91$, $\rho_{SU} = -0.62$, and $\rho_{TU} = -0.83$.
To constrain the model parameter space, we construct the $\chi^2$ function as
\begin{equation}
	\chi^{2}
	=
	\begin{pmatrix}
		\dfrac{S_{\rm th}-S_{\rm exp}}{\sigma_S} &
		\dfrac{T_{\rm th}-T_{\rm exp}}{\sigma_T} &
		\dfrac{U_{\rm th}-U_{\rm exp}}{\sigma_U}
	\end{pmatrix}
	\begin{pmatrix}
    1 & \rho_{ST} & \rho_{SU} \\
    \rho_{ST} & 1 & \rho_{TU} \\
    \rho_{SU} & \rho_{TU} & 1
    \end{pmatrix}^{-1}
	\begin{pmatrix}
		\dfrac{S_{\rm th}-S_{\rm exp}}{\sigma_S} \\[.5em]
		\dfrac{T_{\rm th}-T_{\rm exp}}{\sigma_T} \\[.5em]
		\dfrac{U_{\rm th}-U_{\rm exp}}{\sigma_U}
	\end{pmatrix},
\end{equation}
where $\sigma_S$, $\sigma_T$, and $\sigma_U$ denote the $1\sigma$ uncertainties given in Eq.~\eqref{eq:STU:exp}.
We therefore derive the $95\%$ CL exclusion limits in the $m_{\tau'}$-$s_L$ plane, as shown in Fig.~\ref{fig:chi2}. 
For $m_{\tau'} = 100~(1000)~\si{GeV}$, $s_L \gtrsim 0.42~(0.23)$ has been excluded.

\section{LHC Searches}
\label{sec:simulation}

The primary production processes of the $\tau'$ lepton at the LHC include pair production $p p \to \tau^{\prime +} \tau^{\prime -}$ and single production $p p \to \tau^{\prime \pm} \tau^{\mp}$.
Most existing studies have focused on the pair production process~\cite{CMS:2022cpe,Mahmoud:2024sby,Yue:2024sds}.
The kinematic thresholds for the pair and single production processes are determined by $2m_{\tau'}$ and $m_{\tau'} + m_\tau$, respectively.
The higher threshold for pair production requires relatively large collision energies, suppressing its cross section.
In contrast, the lower threshold of single production enables the LHC to probe a broader region of the parameter space.

\begin{figure}[!t]
\centering
\includegraphics[width=0.5\textwidth]{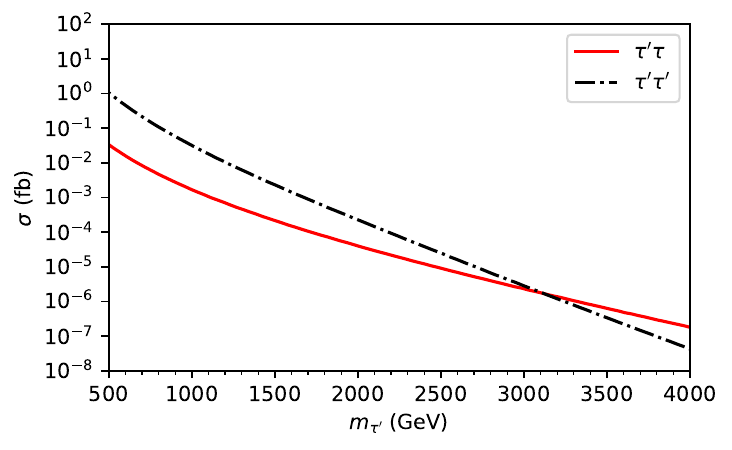}
\caption{Production cross sections for single and pair production of the $\tau'$ lepton as functions of $m_{\tau^\prime}$ with $s_L = 0.1$ at $\sqrt{s} = 14$~TeV.}
\label{fig:cross os}
\end{figure}

The cross sections for single and pair production as functions of $m_{\tau'}$ with $s_L = 0.1$ at $\sqrt{s} = 14$~TeV are computed using \textsc{easyscan}~\cite{Conte:2012fm}, as shown in Fig.~\ref{fig:cross os}.
As the $\tau'$ mass increases, the single production cross section decreases more slowly than the pair production one, owing to its lower kinematic threshold.
Consequently, single production becomes the dominant channel at $m_{\tau'} \gtrsim 3~\si{TeV}$.
Besides the difference in cross section, single production also leads to distinct final states compared to pair production.
This work aims to investigate the sensitivity to the single production of the $\tau'$ lepton at the LHC.

\begin{figure}[t!]
    \centering
    \includegraphics[width=0.5\linewidth]{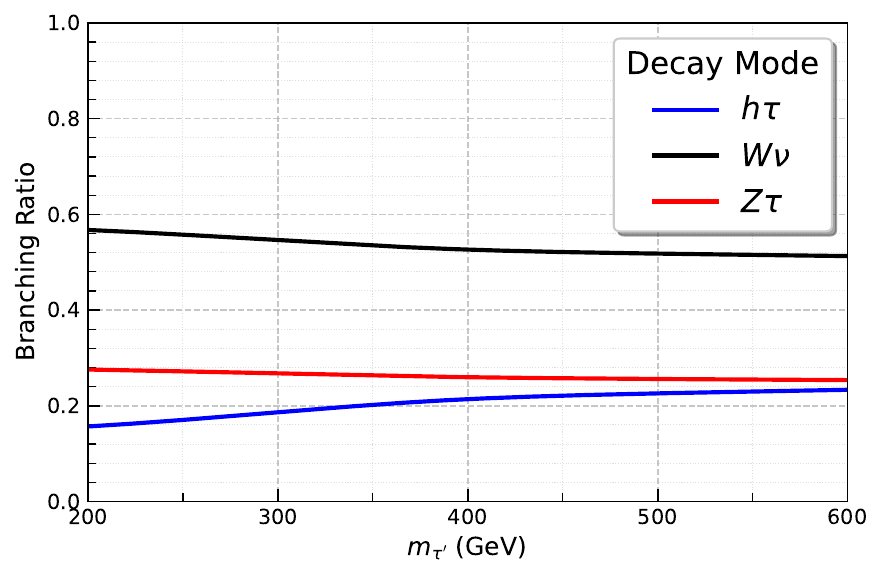}
    \caption{Branching ratios of $\tau^{\prime}$ decay channels for $s_L = 0.1$.}
    \label{fig:branching_ratios}
\end{figure}

For $m_{\tau'} > m_h + m_\tau$, the two-body decay channels of the $\tau'$ lepton include $W \nu_\tau$, $Z\tau$, and $h\tau$, whose branching ratios for $s_L = 0.1$ are displayed in Fig.~\ref{fig:branching_ratios}.
If the singly produced $\tau'$ decays via $W \nu_\tau$, with the $W$ boson decaying leptonically as $W \to \ell \nu_\ell$ ($\ell = e, \mu$), the final state contains only one or two leptons.
Although the $W \nu_\tau$ channel has the largest branching ratio, the resulting low lepton multiplicity provides limited discriminating power for machine learning methods.
Moreover, the $h\tau$ decay channel has the smallest branching ratio and involves the Higgs boson $h$, which predominantly decays to $b\bar{b}$, a final state that is difficult to reconstruct cleanly.
For these reasons, in the subsequent analysis, we concentrate on the $Z\tau$ decay channel, which yields distinct multilepton final states, thereby improving both signal discrimination and reconstruction efficiency.

\begin{figure}[t!]
    \centering
    \includegraphics[width=.36\linewidth]{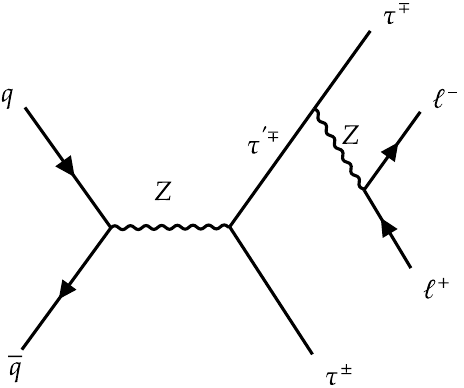}
    \caption{Feynman diagram for the single production of a $\tau^\prime$ lepton via a $Z$ mediator, followed by $\tau^\prime \to Z(\to \ell^+\ell^-) \tau$.}
    \label{fig:Feynman diagram}
\end{figure}

We consider the scenario in which the produced $\tau'$ subsequently decays via $\tau^\prime \to Z\tau$, with the $Z$ boson decaying into $\ell^+ \ell^-$.
The dominant Feynman diagram via a $Z$ mediator is shown in Fig.~\ref{fig:Feynman diagram}.
The relevant $Z\tau'\tau$ coupling originates from the last term in Eq.~\eqref{eq:L_gauge}, whose coefficient is proportional to $c_L s_L$.
Consequently, the single production of $\tau'$ is directly tied to the left-handed $\tau$-$\tau'$ mixing angle $\theta_L$, or equivalently, to the Yukawa coupling $\varepsilon$.

To evaluate the LHC sensitivity to this process, we adopt the following Monte Carlo simulation workflow.
Hard scattering events are generated using \textsc{MadGraph}~\cite{Alwall:2014hca}, with parton distribution functions taken from the \textsc{NN23LO1} set~\cite{Ball:2012cx,Ball:2013hta}.
Particle decays, parton showers, and hadronization are simulated with \textsc{Pythia~8.2}~\cite{Sjostrand:2014zea}, while detector effects are incorporated via \textsc{Delphes~3.5.0}~\cite{deFavereau:2013fsa}. 
Jets are reconstructed using the anti-$k_t$ algorithm~\cite{Cacciari:2008gp} with a radius parameter $R = 0.5$ and a minimum transverse momentum requirement of $p_{\mathrm{T}} > 25~\mathrm{GeV}$.

Depending on the number of the $e$ and $\mu$ leptons in the final state, we classify the search channels of the single $\tau'$ production as the three-lepton and four-lepton channels.
In the three-lepton channel, the final state in Fig.~\ref{fig:Feynman diagram} contains a hadronically decaying $\tau$ ($\tau \to j \nu_{\tau}$) and a leptonically decaying $\tau$ ($\tau \to e\nu_e \nu_\tau$ or $\tau \to \mu\nu_\mu \nu_\tau$).
In the four-lepton channel, both $\tau$ leptons decay leptonically.
Note that the branching ratio of $\tau$ decays into hadrons is $64.8\%$, while that of $\tau$ decays into leptons is $35.2\%$~\cite{ParticleDataGroup:2024cfk}.
In both channels, the $Z$ boson is assumed to decay into $\ell^+\ell^-$.

\subsection{Three-lepton channel}
\label{subsec:trilepton}

In the three-lepton search channel, the major SM backgrounds include double vector boson production $pp \to ZZ$ and triple vector boson production $pp \to Z W^+ W^-$ and $pp \to ZZZ$, where all the $Z$ and $W$ bosons decay into leptons.
For clarity, the signal and major background processes are summarized in Table~\ref{tab:signal final}, where $\tau_h$ and $\tau_\ell$ denote hadronically and leptonically decaying $\tau$ leptons, respectively.

\begin{table}[t!]
\centering
\setlength{\tabcolsep}{1em}
\renewcommand{\arraystretch}{1.5}
\begin{tabular}{ccc}
\toprule[2pt]
    & Production &  Decays \\
\midrule[1pt]
Signal & $pp \to \tau^{\prime \pm} \tau^{\mp}_{h/\ell}$ & 
$\tau^{\prime} \to Z (\to \ell^+ \ell^-) \tau_{\ell/h}$\\
\midrule
& $pp \to ZZ$ &
$Z \to \tau^\pm_h \tau^\mp_\ell$, $Z \to \ell^{+} \ell^{-}$ \\
SM backgrounds & $pp \to ZW^+W^-$ & 
$W^+ \to \ell^+ \nu_\ell$, $W^- \to \ell^- \bar{\nu}_\ell$, $Z \to \tau^\pm_h \tau^\mp_\ell$ \\
& $pp \to ZZZ$ &
$Z \to \nu_\ell \bar{\nu}_\ell$, $Z \to \ell^+ \ell^-$, $Z \to \tau^\pm_h\tau^\mp_\ell$ \\
\bottomrule[2pt]
\end{tabular}
\caption{Signal and major SM background processes in the three-lepton channel. $\tau_h$ and $\tau_\ell$ indicate hadronically and leptonically decaying $\tau$ leptons, respectively.}
\label{tab:signal final}
\end{table}

\begin{figure}[t!]
    \centering
    \begin{subfigure}[b]{0.45\linewidth}
        \includegraphics[width=\linewidth]{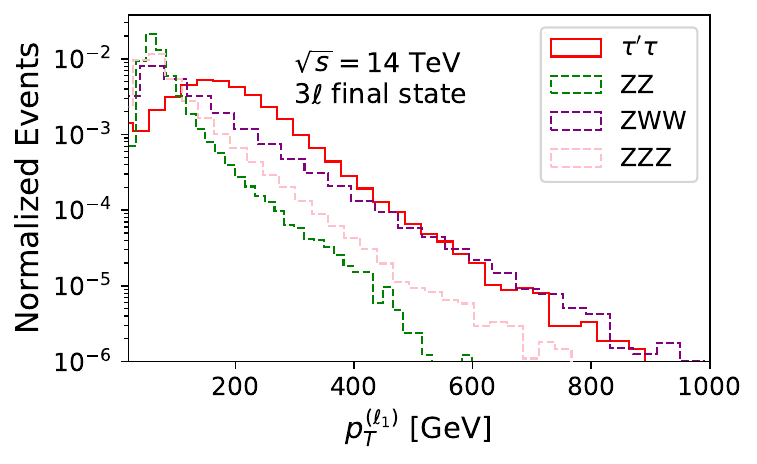}
        \caption{$p_T^{(\ell_1)}$  distribution}
        \label{fig:3l:kinematic:pT_l1}
    \end{subfigure}
    \hspace{1.5em}
    \begin{subfigure}[b]{0.45\linewidth}
        \includegraphics[width=\linewidth]{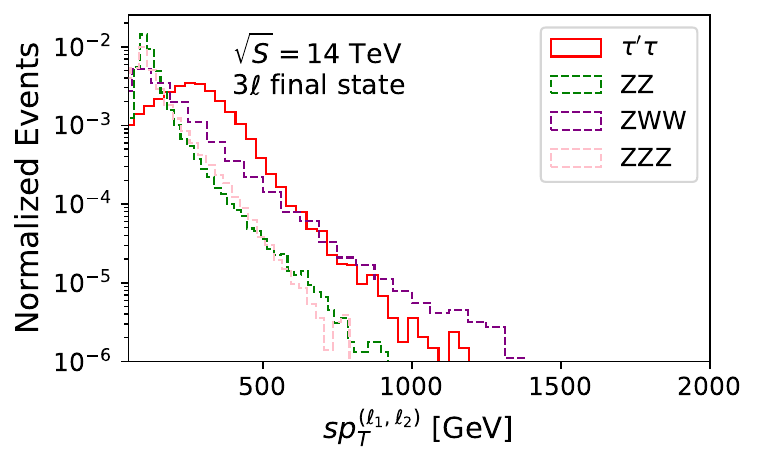}
        \caption{$s p_{T}^{(\ell_1,\ell_2)}$ distribution}
        \label{fig:3l:kinematic:spT}
    \end{subfigure}
    
    \begin{subfigure}[b]{0.45\linewidth}
        \includegraphics[width=\linewidth]{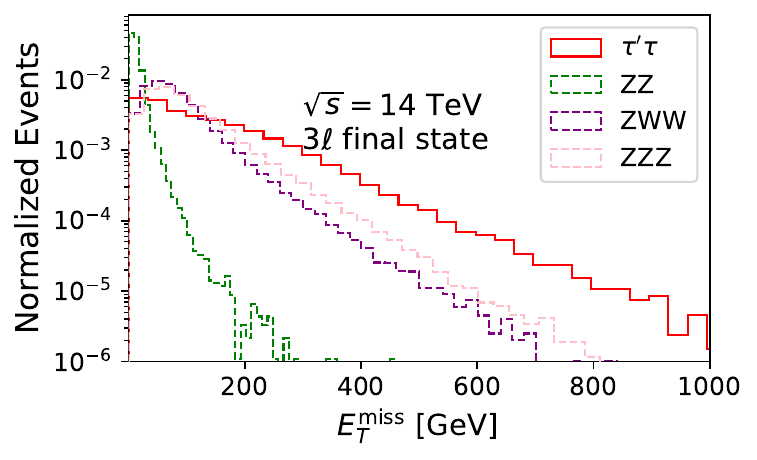}
        \caption{$E_T^{\mathrm{miss}}$ distribution}
        \label{fig:3l:kinematic:ETmiss}
    \end{subfigure}
    \hspace{1.5em}
    \begin{subfigure}[b]{0.45\linewidth}
        \includegraphics[width=\linewidth]{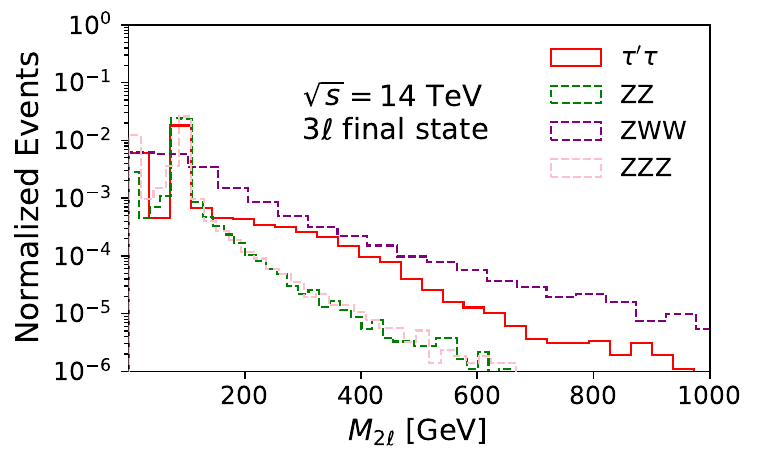}
        \caption{$M_{2\ell}$ distribution}
        \label{fig:3l:kinematic:M_2l}
    \end{subfigure}

    \begin{subfigure}[b]{0.45\linewidth}
        \includegraphics[width=\linewidth]{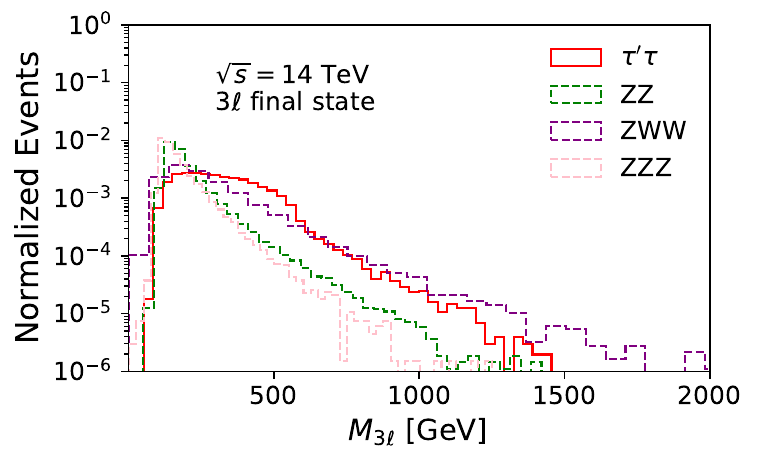}
        \caption{$M_{3\ell}$ distribution}
        \label{fig:3l:kinematic:M_3l}
    \end{subfigure}
    
    \caption{
        Normalized distributions of $p_T^{(\ell_1)}$ (a), $s p_{T}^{(\ell_1,\ell_2)}$ (b), $E_T^{\mathrm{miss}}$ (c), $M_{2\ell}$ (d), and $M_{3\ell}$ (e)
        for the signal benchmark point ($M_{\tau^\prime}=603$~GeV, $s_L=0.1$) and major backgrounds in the three-lepton channel at $\sqrt{s}=14$~TeV.
    }
    \label{fig:3l:kinematic}
\end{figure}

To effectively distinguish signal events from background events, we consider the following primary kinematic variables: 
\begin{itemize}
\item $p_T^{(\ell_1)}$: the transverse momentum of the leading lepton,
\item $s p_{T}^{(\ell_1, \ell_2)}$: the scalar sum of transverse momenta of two leading leptons,
\item $E_T^{\mathrm{miss}}$: the missing transverse energy, defined as the magnitude of the missing transverse momentum $\mathbf{p}_T^\mathrm{miss}$,
\item $M_{2\ell}$: the invariant mass of two opposite-sign same-flavor leptons whose invariant mass is closest to $m_Z$, and
\item  $M_{3\ell}$: the invariant mass of three leptons, two of which are those forming $M_{2\ell}$, while the third is the lepton with highest $p_T$ among the remaining ones.
\end{itemize}
For demonstrating the distributions of the kinematic variables, we adopt a signal benchmark point with $m_{\tau^\prime} = 603$~GeV and $s_L=0.1$.
Normalized distributions of these variables for the Monte Carlo events of the signal and major backgrounds at the LHC with $\sqrt{s} = 14$~TeV are shown in Fig.~\ref{fig:3l:kinematic}.

The transverse momentum of the leading lepton,  $ p_T^{(\ell_1)} $, provides good discrimination between signal and background, as displayed in Fig.~\ref{fig:3l:kinematic:pT_l1}.
In the signal process  $ pp \to \tau^{\prime} \tau$ , the heavy  $ \tau^\prime $ decays to a  $ Z $  boson and a  $ \tau $  lepton, which could subsequently decay to charged leptons.
Because of the large  $ \tau^\prime $  mass, its decay products are highly boosted, yielding leptons with high transverse momenta.  
In contrast, the dominant backgrounds  $ZZ$, $ZWW$, and $ZZZ$ involve $Z$ and $W$ bosons as intermediate states, which are lighter than $\tau'$, leading to softer  $ p_T^{(\ell_1)} $  spectra that peak at low values.
The scalar sum  $ s p_{T}^{(\ell_1, \ell_2)} $ shows a similar behavior, as demonstrated in Fig.~\ref{fig:3l:kinematic:spT}.

The missing transverse energy $E_T^{\mathrm{miss}}$, whose distributions are presented in Fig.~\ref{fig:3l:kinematic:ETmiss}, is primarily contributed by neutrinos, which cannot be measured by detectors at the LHC.  
Each signal event contains multiple neutrinos: one from  $ \tau_h \to j\,\nu_{\tau} $  and two from  $ \tau_\ell \to \ell \nu_{\ell} \nu_{\tau}  $, yielding on average three neutrinos per event.  
Because of the large $ \tau^\prime $ mass, these neutrinos carry significant transverse momenta, leading to large  $ E_T^{\mathrm{miss}} $. 
Additionally, the boost from the heavy parent enhances the transverse momenta of all decay products, leading to a broad distribution with a high-$E_T^{\mathrm{miss}}$ tail.
In contrast, background processes $ZWW$ and $ZZZ$ can also yield multiple neutrinos via  $W$ and $Z$  decays, but their  $ E_T^{\mathrm{miss}} $  spectra are much softer.
This is because the masses of $W$ and $Z$ bosons limit the available phase space.
The $Z Z \to \ell^+\ell^-\ell^+\ell^-$ background produces no genuine neutrinos in the final state, resulting in the softest $ E_T^{\mathrm{miss}} $ distribution among all signal and background processes.

In the  signal process $ pp \to \tau\tau'$, $\tau'$ decays to  $ \tau' \to Z\tau $, where the  $ Z $  boson subsequently decays to a pair of opposite-sign same-flavor leptons $\ell^+\ell^-$.
Thus, the dilepton invariant mass $M_{2\ell}$ is expected to peak around $m_Z = 91.2$~GeV.
Consequently, the $M_{2\ell}$ distribution of the signal in Fig.~\ref{fig:3l:kinematic:M_2l} shows a clear $Z$ peak.
The background processes $ZZ$, $ZWW$, and $ZZZ$ also contains opposite-sign same-flavor leptons $\ell^+\ell^-$ from $Z$ decays, giving arise to similar $Z$ peaks.
Nonetheless, the three-lepton channel contains at least three charged leptons in the final states, which allows for multiple combinations of opposite-sign same-flavor lepton pairs. These mismatches, together with detector effects,  result in broader $M_{2\ell}$ distributions.

By construction, the trilepton invariant mass $M_{3\ell} $ is designed to capture all three charged leptons from the $\tau'$ decay.
As a result, Fig.~\ref{fig:3l:kinematic:M_3l} exhibits a pronounced high-mass tail for the signal, reflecting the large $ \tau^\prime $ mass.  
Because neutrinos from $\tau$ decays are not detected,
the reconstructed mass $M_{3\ell} $ is typically lower than the true  $ \tau^\prime $  mass.
In contrast, SM backgrounds $ZZ$, $ZWW$,
and $ZZZ$ drop rapidly at high $M_{3\ell} $.

In addition to the primary kinematic variables $p_T^{(\ell_1)}$, $s p_{T}^{(\ell_1,\ell_2)}$, $E_T^{\mathrm{miss}}$, $M_{2\ell}$, and $M_{3\ell}$, we also incorporate the following secondary kinematic variables to further enhance the discriminating power:
\begin{itemize}
    \item The transverse momenta, pseudorapidities, and azimuthal angles of the leading and subleading leptons and the leading jet,
    \item the pseudorapidity and azimuthal angle of the missing transverse momentum $\mathbf{p}_T^\mathrm{miss}$, and
    \item the scalar sum of the transverse momenta of all jets.
\end{itemize}
The combination of all the primary and secondary kinematic variables provides substantial discriminating power, enabling effective separation of the signal from SM backgrounds through machine learning methods, which are detailed in Sec.~\ref{sec:ml}.

\subsection{Four-lepton channel}
\label{subsec:fourlepton}

In the four-lepton search channel, the major SM backgrounds involve $pp \to ZZ$, $pp \to Z W^+ W^-$, and $pp \to ZZZ$, where all the $Z$ and $W$ boson decays into light leptons $\ell$ and $\nu_\ell$.
The signal and background processes for this channel are summarized in Table~\ref{tab:signal final1}.

\begin{table}[t!]
\centering
\setlength{\tabcolsep}{1em}
\renewcommand{\arraystretch}{1.5}
\begin{tabular}{ccc}
\toprule[2pt]
    & Production & Decays \\
\midrule[1pt]
Signal & $pp \to \tau^{\prime\pm} \tau^{\mp}_\ell$ & 
$\tau_\ell \to \ell \nu_\ell \nu_{\tau}$, $\tau^\prime \to Z(\to \ell^+ \ell^-) \tau_\ell$ \\
\midrule
 & $pp \to ZZ$ & $Z \to \ell^{+} \ell^{-}$, $Z \to \ell^{+} \ell^{-}$ \\
SM backgrounds & $pp \to ZW^+W^-$ & $Z \to \ell^+ \ell^-$, $W^+ \to \ell^+ \nu_\ell$, $W^- \to \ell^- \bar{\nu}_\ell$ \\
& $pp \to ZZZ$ & $Z \to \ell^+ \ell^-$, $Z \to \ell^+ \ell^-$, $Z \to \nu_\ell \bar{\nu}_\ell$ \\
\bottomrule[2pt]
\end{tabular}
\caption{Signal and major SM background processes in the four-lepton channel.}
\label{tab:signal final1}
\end{table}

In addition to $p_T^{(\ell_1)}$, $s p_{T}^{(\ell_1,\ell_2)}$, $E_T^{\mathrm{miss}}$, and $M_{2\ell}$, whose definitions are the same as those used in the three-lepton channel, we also include the four-lepton invariant mass $M_{4\ell}$ as a primary kinematic variable for the four-lepton channel.
By construction, $M_{4\ell}$ is formed from the two leptons that constitute $M_{2\ell}$ together with an additional opposite-sign same-flavor lepton pair, selected such that $M_{4\ell}$ is maximized.

\begin{figure}[t!]
    \centering
    \begin{subfigure}[b]{0.45\linewidth}
        \includegraphics[width=\linewidth]{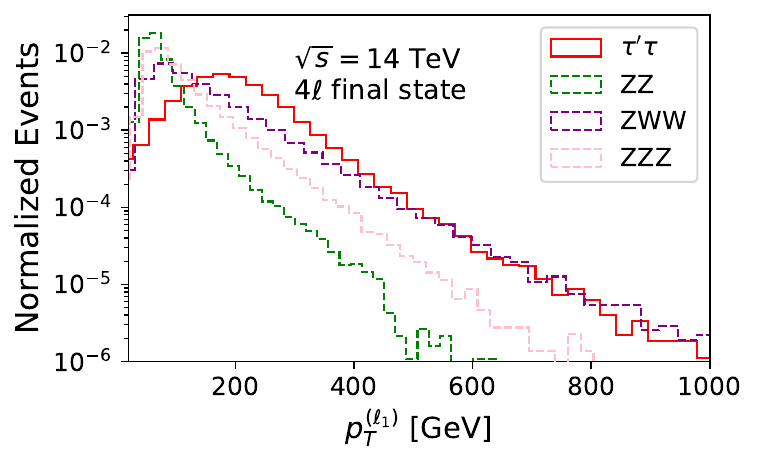}
        \caption{$p_T^{(\ell_1)}$ distribution}
        \label{fig:4l:kinematic:pT_l1}
    \end{subfigure}%
    \hspace{1.5em}
    \begin{subfigure}[b]{0.45\linewidth}
        \includegraphics[width=\linewidth]{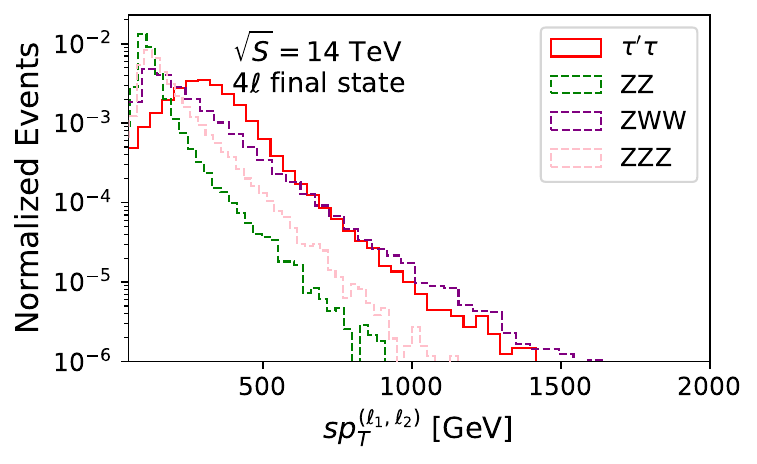}
        \caption{$s p_{T}^{(\ell_1,\ell_2)}$ distribution}
        \label{fig:4l:kinematic:spT}
    \end{subfigure}
    
    \begin{subfigure}[b]{0.45\linewidth}
        \includegraphics[width=\linewidth]{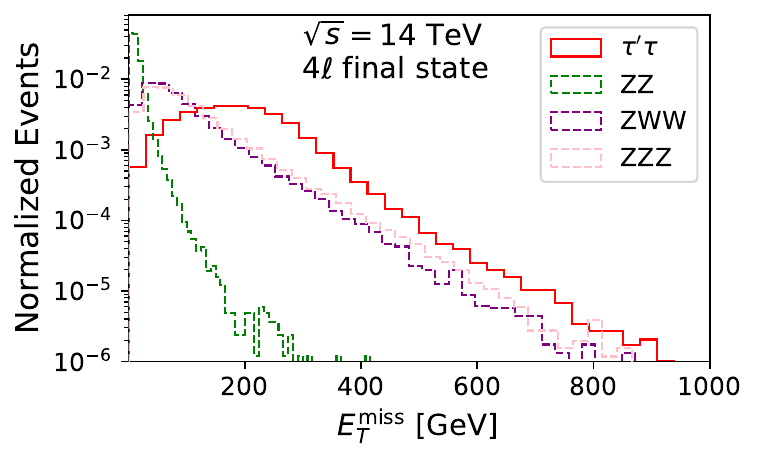}
        \caption{$E_T^{\mathrm{miss}}$ distribution}
        \label{fig:4l:kinematic:ETmiss}
    \end{subfigure}%
    \hspace{1.5em}
    \begin{subfigure}[b]{0.45\linewidth}
        \includegraphics[width=\linewidth]{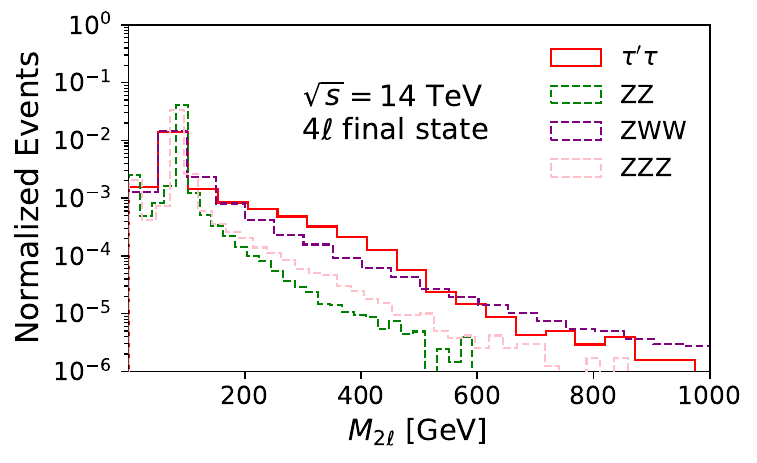}
        \caption{$M_{2\ell}$ distribution}
        \label{fig:4l:kinematic:M_2l}
    \end{subfigure}

    \begin{subfigure}[b]{0.45\linewidth}
        \includegraphics[width=\linewidth]{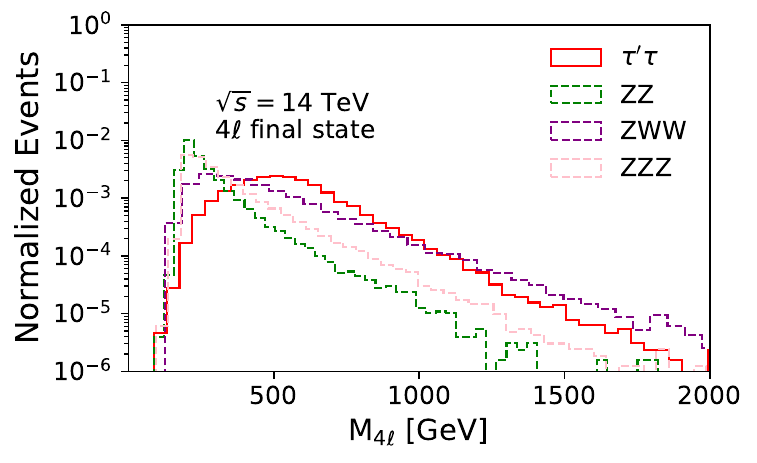}
        \caption{$M_{4\ell}$ distribution}
        \label{fig:4l:kinematic:M_4l}
    \end{subfigure}
    
    \caption{
        Normalized distributions of $p_T^{(\ell_1)}$ (a), $s p_{T}^{(\ell_1,\ell_2)}$ (b), $E_T^{\mathrm{miss}}$ (c), $M_{2\ell}$ (d), and $M_{4\ell}$ (e)
        for the signal benchmark point ($M_{\tau^\prime}=603$~GeV, $s_L = 0.1$) and major backgrounds in the four-lepton channel at $\sqrt{s}=14$~TeV.
    }
    \label{fig:4l:kinematic}
\end{figure}

Figure~\ref{fig:4l:kinematic} shows the normalized distributions of these kinematic variables at $\sqrt{s} = 14$~TeV for the backgrounds and the signal with $M_{\tau^\prime} = 603$~GeV ad  $s_L=0.1$. 
The $p_T^{(\ell_1)}$, $s p_{T}^{(\ell_1, \ell_2)}$, and $M_{2\ell}$ distributions in Figs.~\ref{fig:4l:kinematic:pT_l1}, \ref{fig:4l:kinematic:spT}, and \ref{fig:4l:kinematic:M_2l} are quite similar to those in the three-lepton channel.

The dominant background, $pp \to ZZ$, is expected to produce no genuine missing transverse energy, as all final-state leptons from $Z$ decays are detectable.
Thus, the observed $E_T^{\mathrm{miss}}$ in the $ZZ$ background arises primarily from detector effects, leading to a distribution that falls rapidly from low $E_T^{\mathrm{miss}}$, as shown in Fig.~\ref{fig:4l:kinematic:ETmiss}.
In contrast, the signal process involves four neutrinos from leptonically decays of two $\tau$ leptons, resulting in a harder and broader $E_T^{\mathrm{miss}}$ distribution.
This feature offers strong discrimination between the signal and the SM backgrounds.

The four-lepton invariant mass $M_{4\ell}$ is a crucial observable for reconstructing the $\tau^\prime$ mass.
In the ideal case with negligible neutrino momenta, the invariant mass of the four detected leptons originating from the decay chain  
 $\tau^{\prime} \to Z \tau \to \ell^+\ell^-\ell \nu_\ell \nu_\tau$ and $ \tau \to \ell \nu_\ell \nu_\tau$ would be close to the invariant mass of the $\tau'\tau$ system, whose lower bound is $m_{\tau'} + m_\tau$.
However, because of undetected neutrinos, the reconstructed  $M_{4\ell}$ is typically lower than the ideal case.
For the benchmark point with $m_{\tau'} = 603~\si{GeV}$, the signal $M_{4\ell}$ distribution peaks around $500~\si{GeV}$, as demonstrated in Fig.~\ref{fig:4l:kinematic:M_4l}.
On the other hand, the $M_{4\ell}$ distribution of the SM backgrounds peaks between $200$ and $400~\si{GeV}$, providing great discrimination power.

In addition to the primary kinematic variables, we also include secondary variables in the four-lepton channel.
These are generally similar to those used in the three-lepton channel, although jet-related variables are excluded.

\subsection{Cut-and-count analysis}

Before applying machine learning methods to enhance signal‑background discrimination, we first discuss the conventional cut‑and‑count approach in this subsection.
The results obtained here serve as a baseline for comparison with those presented in the following section, illustrating the improvement achieved with machine learning techniques.

We now apply appropriate kinematic selection cuts to the three- and four-lepton channels.
For the three-lepton channel, we require $p_T^{(\ell_1)} > 100~\si{GeV}$, $sp_T^{(\ell_1, \ell_2)} > 250~\si{GeV}$, $M_{3\ell} > 250~\si{GeV}$, $30~\si{GeV} <M_{2\ell} < 200~\si{GeV}$, and $E_T^{\mathrm{miss}} > 50~\si{GeV}$. 
For the four-lepton channel, the selection criteria are $p_T^{(\ell_1)} > 100~\si{GeV}$, $sp_T^{(\ell_1, \ell_2)} > 250~\si{GeV}$, $M_{4\ell} > 300~\si{GeV}$, $30~\si{GeV} <M_{2\ell} < 200~\si{GeV}$, and $E_T^{\mathrm{miss}} > 80~\si{GeV}$.

The selection efficiencies after applying all kinematic cuts in both channels are summarized in Table~\ref{table:cut_efficiencies}.
In the three-lepton (four-lepton) channel, the selection efficiency for the dominant background $pp \to ZZ$ is $0.13\%$ ($0.017\%$), whereas the signal efficiency reaches about $16\%$ ($21\%$).

\begin{table}[!t]
\centering
\setlength{\tabcolsep}{1em}
\renewcommand{\arraystretch}{1.5}
\begin{tabularx}{.9\textwidth}{
    @{\hspace{3em}} 
    >{\raggedright\arraybackslash}X 
    C
    >{\raggedright\arraybackslash}X 
    C
}
  \toprule[2pt]
  \multicolumn{2}{@{\hspace{-\tabcolsep}}c}{\textbf{Three-lepton channel}}
  & \multicolumn{2}{c}{\textbf{Four-lepton channel}} \\
  \cmidrule(lr){1-2} \cmidrule(l){3-4}
  \textbf{Process} & \textbf{Efficiency (\%)} 
  & \textbf{Process} & \textbf{Efficiency (\%)} \\
  \midrule[1pt]
  $pp \to \tau^{\prime \pm} \tau^{\mp}_h$ & 15.873 
  & $pp \to \tau^{\prime \pm} \tau^{\mp}_\ell$ & 21.050   \\
  $pp \to ZZ$                       & 0.130 
  & $pp \to ZZ$                       & 0.017  \\
  $pp \to ZW^+W^-$                  & 4.288 
  & $pp \to ZW^+W^-$                  & 4.348   \\
  $pp \to ZZZ$                      & 1.432
  & $pp \to ZZZ$                      & 5.177   \\
  \bottomrule[2pt]
\end{tabularx}
\caption{Signal and background efficiencies after kinematic selection cuts for the three-lepton and four-lepton channels.}
\label{table:cut_efficiencies}
\end{table}

\section{Machine learning analysis}
\label{sec:ml}

In this section, we employ the machine learning algorithm XGBoost to improve the discrimination between signal and background events.
XGBoost (eXtreme Gradient Boosting)~\cite{Chen:2016btl} is an enhanced ensemble learning method based on gradient boosting decision trees (GBDT)~\cite{Friedman:2001wbq}. 
Compared to the standard GBDT, 
XGBoost uses both first and second order derivatives of the loss function, allowing a more accurate approximation of the objective function.
It also includes a regularization term to control model complexity and reduce overfitting.
At each node of a decision tree, the algorithm evaluates the gain for all possible splits and selects the one with the highest gain to optimize classification performance.
The learning rate is tuned to optimize training,
making XGBoost one of the most widely used and effective machine learning methods.

\subsection{Model training}
\label{subsec:training}

\begin{figure}[t!]
    \centering
    \begin{subfigure}[b]{0.45\linewidth}
        \includegraphics[width=\linewidth]{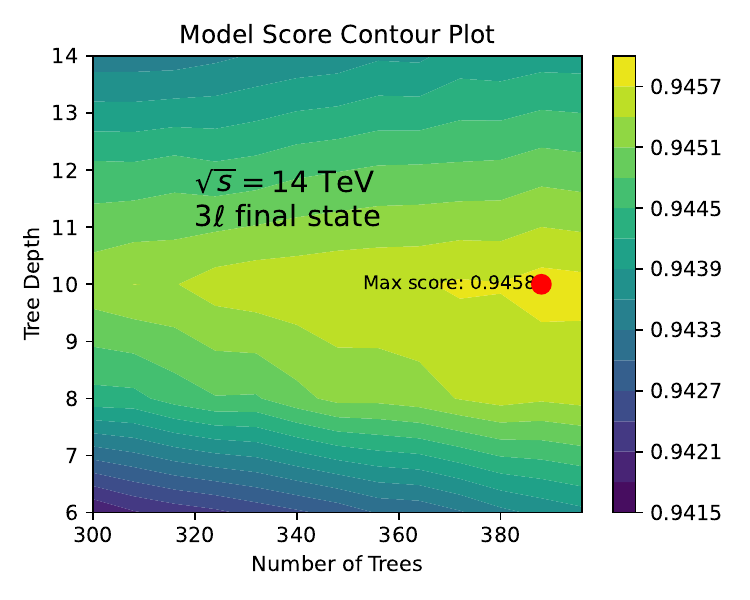}
        \caption{Three-lepton channel}
    \end{subfigure}%
    \hspace{1.5em}
    \begin{subfigure}[b]{0.45\linewidth}
        \includegraphics[width=\linewidth]{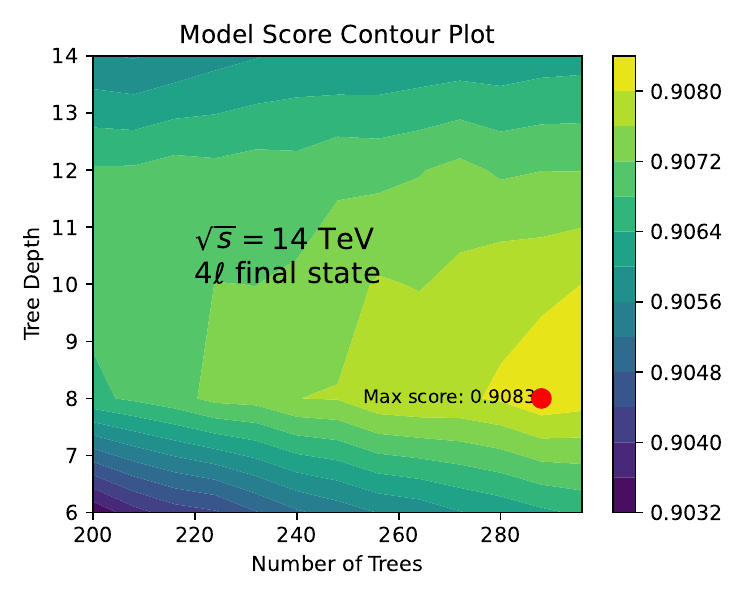}
        \caption{Four-lepton channel}
    \end{subfigure}
    \caption{Hyperparameter optimization scores for the three-lepton (left) and four-lepton (right) channels.}
    \label{fig:hyperparameter score}
\end{figure}

The XGBoost model is trained using the kinematic variables in the three- and four-lepton channels described in the previous sections.  
The key hyperparameters employed in the training are listed below.
\begin{itemize}
    \item Learning rate: this controls the step size per iteration. A smaller value improves stability but requires more training rounds. We set it to 0.01.
    \item Maximum depth (\texttt{max\_depth}): this limits the depth of each decision tree to prevent overfitting.
    \item Regularization parameters: we set them as the default XGBoost values to control model complexity.
    \item Number of trees (\texttt{n\_estimators}): this is the total number of boosting rounds. Too many trees may lead to overfitting.
\end{itemize}
The performance scores for different hyperparameter settings in both channels are shown in Fig.~\ref{fig:hyperparameter score}. 
After optimization, we adopt \texttt{max\_depth = 10} and 
\texttt{n\_estimators = 388} for the three-lepton channel, and \texttt{max\_depth = 8} and \texttt{n\_estimators = 288}) for the four-lepton channel.
With these settings, the model is trained on datasets simulated with \textsc{MadGraph}.

As illustrated in Fig.~\ref{fig:hyperparameter score}, the optimal hyperparameter regions for the two channels are quite different. Specifically, due to the complex kinematic features of the three-lepton channel, the model requires a higher computational complexity to fully capture the distinctions between the signal and backgrounds. Consequently, the high-score regions in the performance heatmap correspond to a larger number of decision trees and greater tree depths. In contrast, the classification task for the four-lepton channel is relatively lighter, with signal features that are more easily separable in high-dimensional space, thus requiring fewer trees for convergence. At the 14~TeV LHC, the optimal number of trees and maximum depth for the four-lepton channel are comparatively smaller. This indicates that a lower model capacity is sufficient to extract adequate discriminative information. Furthermore, the optimal score regions in the heatmaps of both channels are concentrated and relatively flat, suggesting that the parameter settings are reasonable and that the model is insensitive to minor hyperparameter variations, yielding robust training results.

\subsection{Sensitivity at the 14~TeV LHC}
\label{subsec:sensitivity}

The sensitivity of the LHC to the VLL $\tau'$ depends on several factors,
including the VLL mass $m_{\tau^\prime}$, the mixing angle $s_L$, the collision energy $\sqrt{s}$, the integrated luminosity $L$, and the detector performance.
The expected numbers of signal and background events are defined as $s = \sigma_{S} L$ and $b = \sigma_{B} L$, where $\sigma_{S}$ and $\sigma_{B}$ are the cross sections of the signal and the total backgrounds after applying the machine learning selection.
Using the profile-likelihood-ratio test statistic, the significance $S$ is given by~\cite{Kumar:2015tna}
\begin{equation}
S = \sqrt{ 2 \left[ (s+b) \ln  \frac{(s+b)(b+\Delta_b^2)}{b^2 + (s+b)\Delta_b^2}  - \frac{b^2}{\Delta_b^2} \ln \left( 1 + \frac{\Delta_b^2 s}{b(b+\Delta_b^2)} \right) \right] },
\label{eq:significances_precies}
\end{equation}
where $\Delta_b$ denotes the systematic uncertainty on the background event number.
When systematic uncertainties are negligible, the expression above reduces to the statistical significance formula based on Poisson statistics~\cite{Cowan:2010js},
\begin{equation}
S = \sqrt{2 \left[ (s+b) \ln \left(1 + \frac{s}{b}\right) - s \right]}.
\end{equation}

\begin{figure}[!t]
    \centering
    \begin{subfigure}[b]{0.45\linewidth}
        \includegraphics[width=\linewidth]{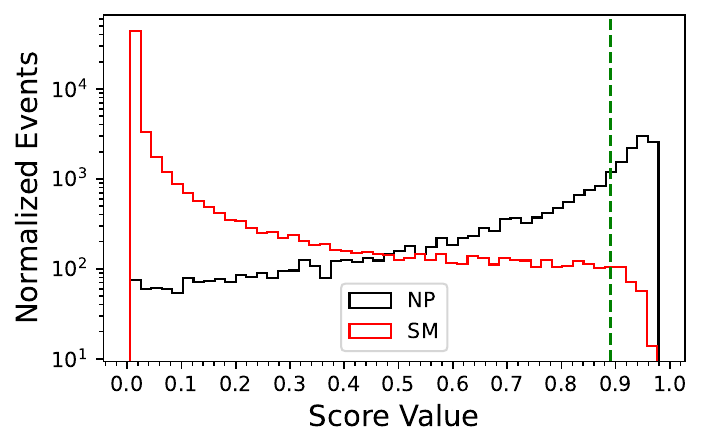}
        \caption{Three-lepton channel}
    \end{subfigure}%
    \hspace{1.5em}
    \begin{subfigure}[b]{0.45\linewidth}
        \includegraphics[width=\linewidth]{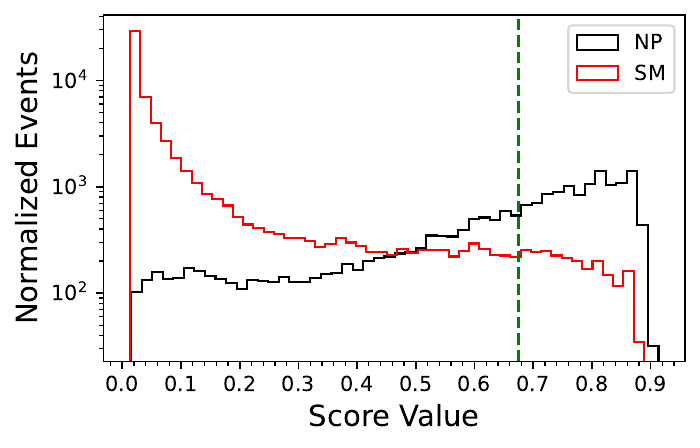}
        \caption{Four-lepton channel}
    \end{subfigure}
    \caption{Distributions of the XGBoost classifier output for the signal (black) and the SM backgrounds (red) in the three-lepton (a) and four-lepton (b) channels. The green dashed line indicates the optimal threshold $\chi$.}
    \label{fig:optimal threshold}
\end{figure}

To further improve classification, 
we scan over all possible output thresholds of the XGBoost classifier and select the one that maximizes the significance $S$, assuming $\Delta_b = 0$, which is a reasonable approximation when statistical uncertainties dominate over systematic uncertainties.
Table~\ref{table:efficiencies} summarizes the selection efficiencies in the three- and four-lepton channels, where $\chi$ represents the optimal threshold.
It shows that the XGBoost algorithm greatly suppresses the backgrounds while keeping a reasonable signal efficiency. 
For the dominant background $pp \to ZZ$, the selection efficiency in the three-lepton channel is $\mathcal{O}(10^{-3})$, while in the four-lepton channel it is $\mathcal{O}(10^{-4})$.
Compared with the cut-and-count results in Table~\ref{table:cut_efficiencies}, the selection efficiency for $pp \to ZZ$ remains similar, while the signal efficiency increases by a factor of $\sim 3.4$ in both channels.
This yields a substantial improvement in the signal significance.
The classifier output distributions are presented in Fig.~\ref{fig:optimal threshold}.
For both channels, the signal peaks near $1$, while the backgrounds peaks near $0$.
The green dashed line indicates the optimal threshold $\chi$.

\begin{table}[!t]
\centering
\setlength{\tabcolsep}{1em}
\renewcommand{\arraystretch}{1.5}
\begin{tabularx}{.9\textwidth}{
    @{\hspace{3em}} 
    >{\raggedright\arraybackslash}X 
    C
    >{\raggedright\arraybackslash}X 
    C
}
  \toprule[2pt]
  \multicolumn{2}{@{\hspace{-\tabcolsep}}c}{\textbf{Three-lepton channel} ($\chi = 0.8740$)}
  & \multicolumn{2}{c}{\textbf{Four-lepton channel} ($\chi = 0.6750$)} \\
  \cmidrule(lr){1-2} \cmidrule(l){3-4}
  \textbf{Process} & \textbf{Efficiency (\%)} 
  & \textbf{Process} & \textbf{Efficiency (\%)} \\
  \midrule[1pt]
  $pp \to \tau^{\prime \pm} \tau^{\mp}_h$ & 54.375 
  & $pp \to \tau^{\prime \pm} \tau^{\mp}_\ell$ & 72.805   \\
  $pp \to ZZ$                       & 0.150 
  & $pp \to ZZ$                       & 0.010  \\
  $pp \to ZW^+W^-$                  & 0.575 
  & $pp \to ZW^+W^-$                  & 5.635   \\
  $pp \to ZZZ$                      & 1.175
  & $pp \to ZZZ$                      & 5.625   \\
  \bottomrule[2pt]
\end{tabularx}
\caption{Signal and background efficiencies after optimized machine learning selection for the three-lepton and four-lepton channels. The $\chi$ values indicate the thresholds used for significance maximization.}
\label{table:efficiencies}
\end{table}

\begin{figure}[!t]
    \centering
    \begin{subfigure}{0.47\linewidth}
        \includegraphics[width=\linewidth]{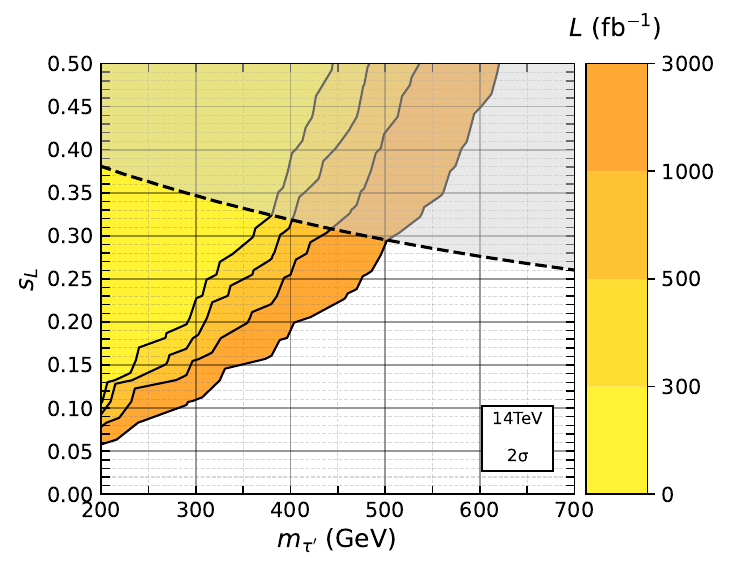}
        \caption{Three-lepton, $2\sigma$  exclusion}
    \end{subfigure}%
    \hspace{1.5em}
    \begin{subfigure}{0.47\linewidth}
        \includegraphics[width=\linewidth]{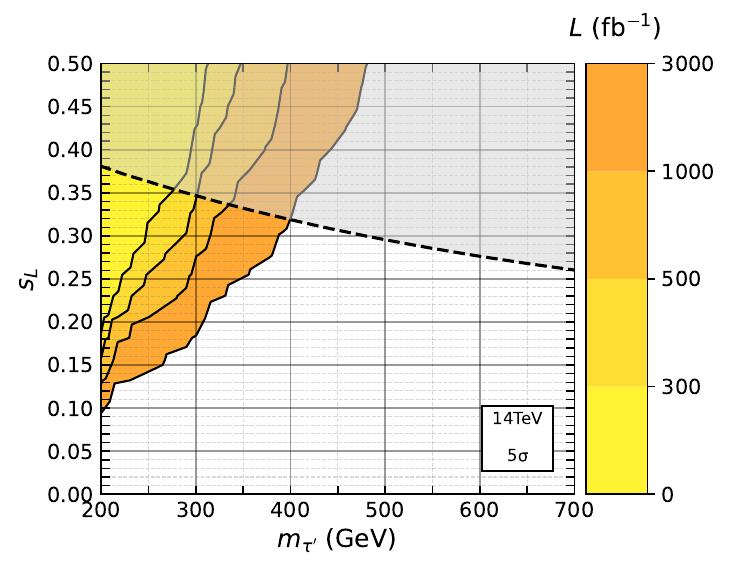}
        \caption{Three-lepton, $5\sigma$ discovery}
    \end{subfigure}

    \begin{subfigure}{0.47\linewidth}
        \includegraphics[width=\linewidth]{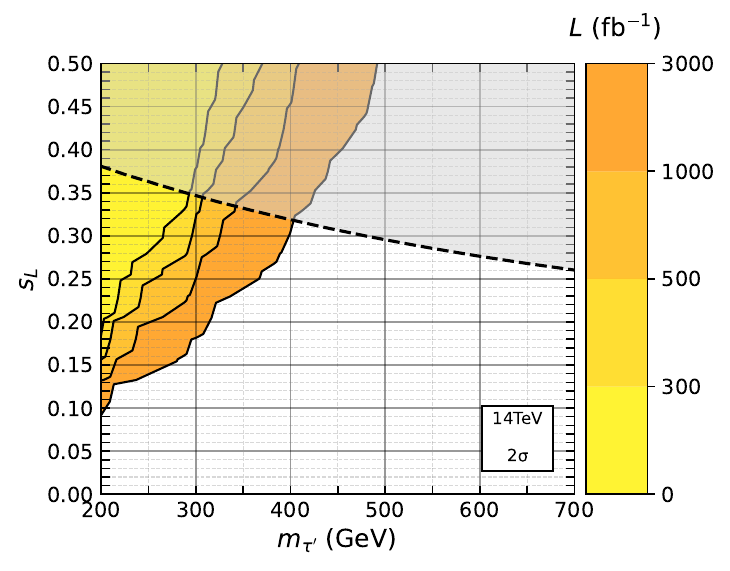}
        \caption{Four-lepton, $2\sigma$ exclusion}
    \end{subfigure}%
    \hspace{1.5em}
    \begin{subfigure}{0.47\linewidth}
        \includegraphics[width=\linewidth]{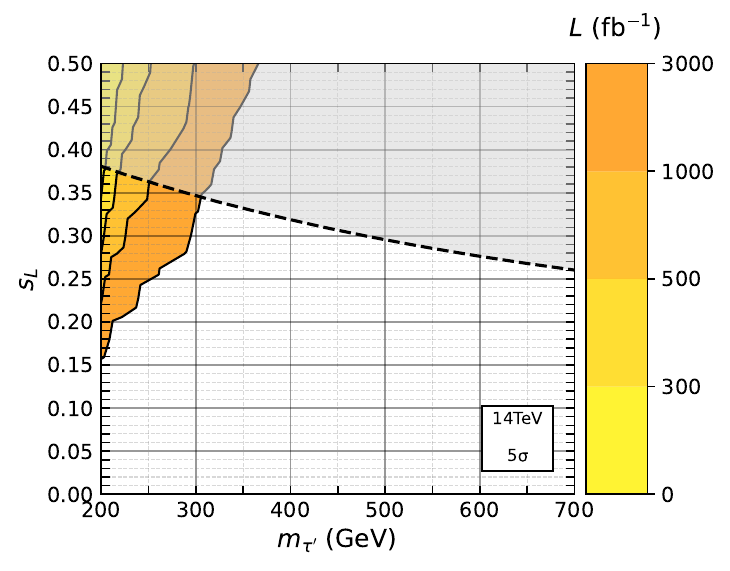}
        \caption{Four-lepton, $5\sigma$ discovery}
    \end{subfigure}

    \caption{Expected $2\sigma$ exclusion (left) and $5\sigma$ discovery (right) limits in the $m_{\tau^\prime}$-$s_L$ plane for the three-lepton (upper) and four-lepton (lower) channels at the $14~\mathrm{TeV}$ LHC with integrated luminosity $L = 300,500,1000,3000~\si{fb^{-1}}$, assuming negligible systematic uncertainties. The black dashed lines indicate the $95\%$ CL exclusion limits derived from electroweak oblique parameters.}
    \label{fig:exclusion}
\end{figure}

We now investigate the LHC sensitivity to the VLL after applying the machine-learning-based selection, under the optimistic assumption that systematic uncertainties are negligible.
Parameter scans are performed in the $m_{\tau^\prime}$-$s_L$ plane using \textsc{easyscan}.
Sensitivity curves are obtained by requiring  $S = 2$  ($2\sigma$ exclusion limit) and  $S = 5$  ($5\sigma$ discovery limit).
Figure~\ref{fig:exclusion} shows the expected $2\sigma$ exclusion and $5\sigma$ discovery limits in the  $m_{\tau^\prime}$-$s_L$  plane for the three- and four-lepton search channels at the LHC with $\sqrt{s} = 14~\mathrm{TeV}$ and integrated luminosity $L = 300,500,1000,3000~\si{fb^{-1}}$.
For comparison, the $95\%$ CL exclusion limits derived from electroweak oblique parameters are also demonstrated as black dashed lines.

For $s_L = 0.3$ in the three-lepton channel with an integrated luminosity of $300~\mathrm{fb}^{-1} $, the expected $2\sigma$ exclusion ($5\sigma$ discovery) limit reaches up to $m_{\tau'} \simeq 360~(245)~\si{GeV}$. 
At the HL-LHC stage with $L =  3000~\mathrm{fb}^{-1} $, the $2\sigma$ exclusion limit can reach the point $(m_{\tau^\prime}, s_L) = (500~\si{GeV}, 0.29) $, and the $5\sigma$ discovery limit improves to $m_{\tau'} \simeq 385~\si{GeV}$ for $s_L = 0.3$.
For $ s_L = 0.3$ in the four-lepton channel, the $2\sigma$ exclusion limit with a dataset of $300~(3000)~\mathrm{fb}^{-1}$ reaches up to $m_{\tau'} \simeq 265~(400)~\si{GeV}$, while the $5\sigma$ discovery limit with $L = 3000~\si{fb^{-1}}$ can probe $m_{\tau'} \lesssim 295~\si{GeV}$.
These results demonstrate that the three-lepton channel at the 14~TeV LHC is more sensitive to single $\tau'$ production than the four-lepton channel, primarily owing to the larger branching ratio of $\tau$ decays into hadrons.

\begin{figure}[!t]
    \centering
    \begin{subfigure}{0.47\linewidth}
        \includegraphics[width=\linewidth]{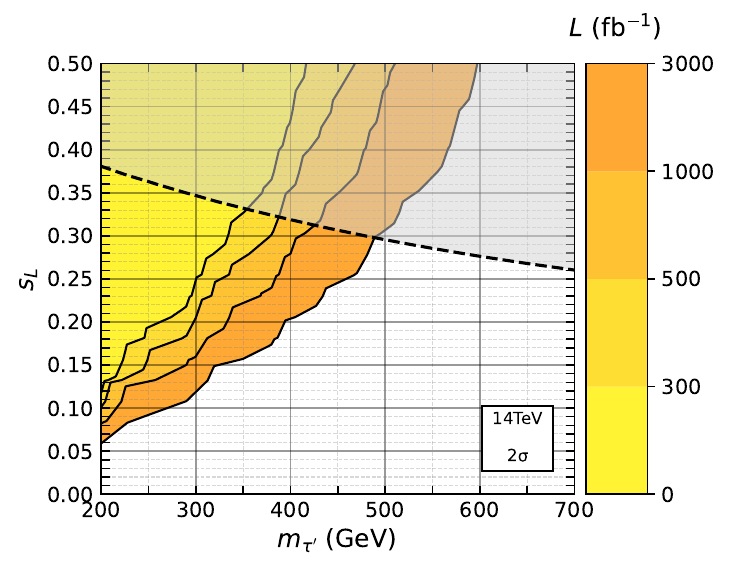}
        \caption{Three-lepton, $2\sigma$  exclusion}
    \end{subfigure}%
    \hspace{1.5em}
    \begin{subfigure}{0.47\linewidth}
        \includegraphics[width=\linewidth]{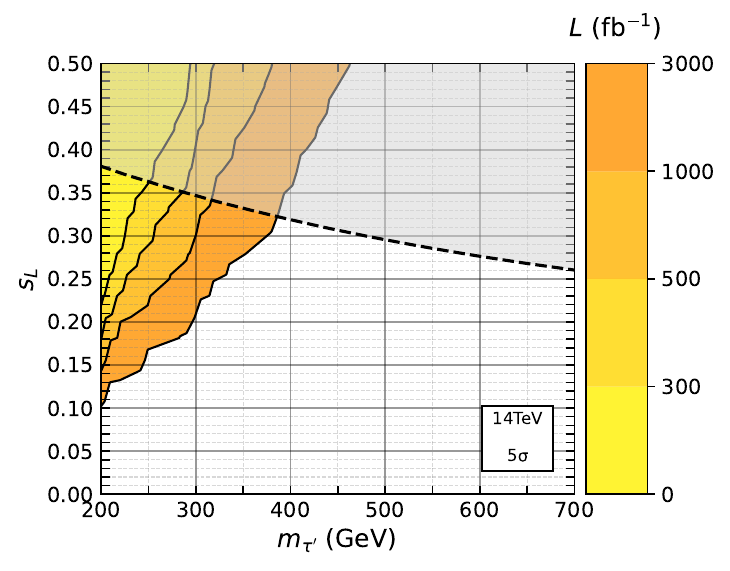}
        \caption{Three-lepton, $5\sigma$ discovery}
    \end{subfigure}

    \begin{subfigure}{0.47\linewidth}
        \includegraphics[width=\linewidth]{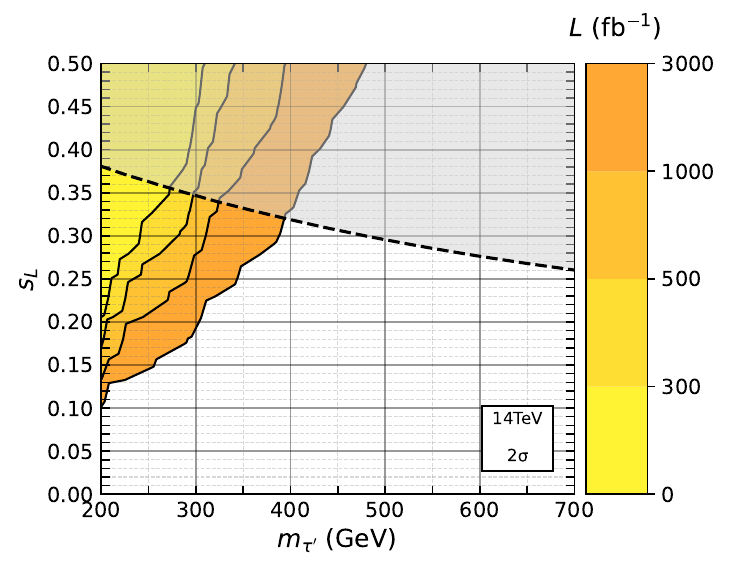}
        \caption{Four-lepton, $2\sigma$ exclusion}
    \end{subfigure}%
    \hspace{1.5em}
    \begin{subfigure}{0.47\linewidth}
        \includegraphics[width=\linewidth]{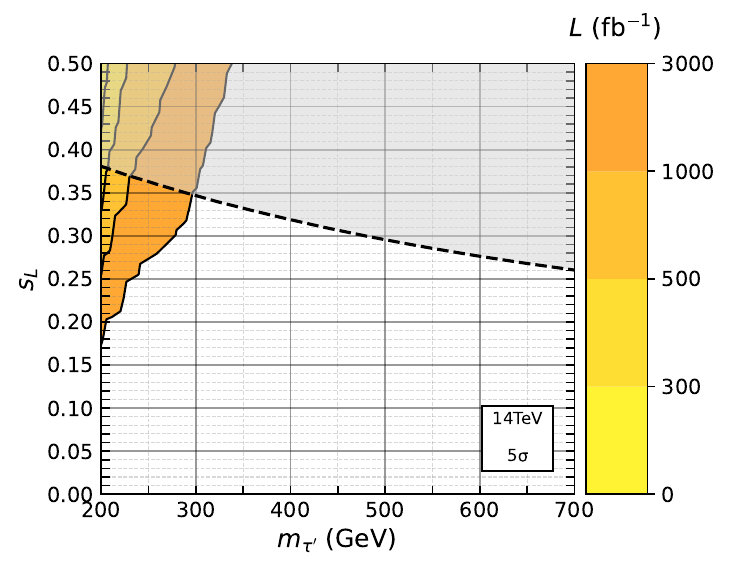}
        \caption{Four-lepton, $5\sigma$ discovery}
    \end{subfigure}

    \caption{Expected $2\sigma$ exclusion (left) and $5\sigma$ discovery (right) limits in the $m_{\tau^\prime}$-$s_L$ plane for the three-lepton (upper) and four-lepton (lower) channels at the $14~\mathrm{TeV}$ LHC with integrated luminosity $L = 300,500,1000,3000~\si{fb^{-1}}$, under the assumption of $\Delta_b = 0.5\sqrt{b}$. The black dashed lines indicate the $95\%$ CL exclusion limits derived from electroweak oblique parameters.}
    \label{fig:exclusion_precise}
\end{figure}

To provide a more realistic sensitivity estimate, we also examine the scenario in which systematic uncertainties are significant.
Accordingly, we set $\Delta_b = 0.5\sqrt{b}$, corresponding to systematic uncertainties at $50\%$ of the statistical uncertainties, and recalculate the significance.
The resulting exclusion and discovery limits are displayed in Fig.~\ref{fig:exclusion_precise}.
For $s_L = 0.3$, the $2\sigma$ exclusion limit with $L = 3000~\si{fb}^{-1}$ reaches up to $m_{\tau'} \simeq 490~(390)~\si{GeV}$ in the three-lepton (four-lepton) channel.

\section{Conclusions}
\label{sec:conclusion}

The exploration and discovery of new particles at colliders is one of the most important goals in particle physics.  
Vectorlike leptons represent a well-motivated extension of the Standard Model.  
In this paper, we focus on the single production of a singlet $\tau'$ lepton, $ pp \to \tau^{\prime\pm} \tau^\mp$ with $\tau^{\prime\pm} \to Z\tau^\pm$, at the LHC and investigate how machine learning techniques can enhance their detection sensitivity. 

Depending on the decay processes of two $\tau$ leptons, the final states are classified into the three- and four-lepton search channels.
In the  three-lepton channel, the primary kinematic variables include the transverse momentum of the leading lepton $ p_{\mathrm{T}}^{(\ell_1)} $, the scalar sum of the transverse momenta of two leading leptons $ s p_T^{(\ell_1, \ell_2)} $, the missing transverse energy  $ E_{T}^{\mathrm{miss}} $, the dilepton invariant mass $ M_{2\ell} $, and the trilepton invariant mass $ M_{3\ell} $.
In the four-lepton channel,  $M_{3\ell} $ is replaced by the four-lepton invariant mass $ M_{4\ell} $.

In order to enhance signal-background discrimination, independent XGBoost classifiers are trained for both channels.
Classifier hyperparameters, such as the number of trees and the maximum tree depth, are optimized via grid search on the training set using average cross-validation scores, including negative log loss and weighted accuracy.  
Notably, within the same hyperparameter search space, the classifier for the four-lepton channel achieves a higher optimal cross-validation score than that for the three-lepton channel, indicating superior model fitting and generalization.  
Consequently, for the dominant SM background $pp \to ZZ$, the selection efficiencies in the three- and four-lepton channels are $\mathcal{O}(10^{-3})$ and $\mathcal{O}(10^{-4})$, respectively.
This improvement stems from stronger kinematic constraints in the four-lepton channel, which enhance the separability between signal and background in the feature space.

After adopting appropriate output thresholds of the XGBoost classifier, the SM backgrounds in both channels can be greatly suppressed, while the signal efficiency remain sufficiently high.
Our analyses show that, at the 14~TeV HL-LHC with an integrated luminosity of $ 3000~\mathrm{fb}^{-1} $ under the assumption of negligible systematic uncertainties, the $2\sigma$ exclusion limit in the three-lepton (four-lepton) channel can reach up to $m_{\tau'} \simeq 500~(405)~\si{GeV}$ in the parameter region allowed by the electroweak oblique parameter constraint.
If the $5\sigma$ discovery limit is considered, the HL-LHC can probe a $\tau'$ lepton with a mass of $\sim 400~(305)~\si{GeV}$ in the three-lepton (four-lepton) channel.
If systematic uncertainties are assumed to be $50\%$ of the statistical uncertainties, the resulting signal significance is only slightly reduced.

\begin{acknowledgments}
		
This work is supported by the National Natural Science Foundation of China (NSFC) under Grants No.~12275367, No.~11875327, and No.~12575115, the Fundamental Research Funds for the Central Universities, and the Sun Yat-Sen University Science Foundation.
		
\end{acknowledgments}

\bibliographystyle{utphys}
	
\bibliography{ref1}

\end{document}